# High performance of carbon nanotube refrigerators over a large temperature span

*Tatiana Naomi Yamamoto Silva, Alexandre F. Fonseca* *

Applied Physics Department, Institute of Physics "Gleb Wataghin", University of Campinas - UNICAMP, 13083-859, Campinas, São Paulo, Brazil.

**Corresponding Author**

* Phone: +55 19 3521-5364. Email: Alexandre F. Fonseca – afonseca@ifi.unicamp.br

**ABSTRACT:** Compression of greenhouse gases still dominates the market of refrigeration devices. Although well stablished and efficient, this technology is neither safe for the environment nor able to be scaled down to nanoscale. Solid-state cooling technologies are being developed to overcome these limitations, including studies at nanoscale. Among them, the so-called elastocaloric effect (eC) consists of the thermal response, $\Delta T$, of a material under strain deformation. In this work, fully atomistic molecular dynamics simulations of the eC in carbon nanotubes (CNTs) are presented over a large temperature span. The efficiency of the CNTs as solid refrigerators is investigated by simulating their eC in a model of refrigerator machine running under Otto-like thermodynamic cycles (two adiabatic expansion/contraction plus two isochoric heat exchange processes) operating at temperatures, $T_O$, ranging from 300 to 2000 K. The coefficient-of-performance (COP), defined as the ratio of heat removed from the cold region to the total work performed by the system per thermodynamic cycle, is calculated for each value of $T_O$. Our results show a non-linear dependence of $\Delta T$ on $T_O$, reaching a minimum value of about 30 K for $T_O$ between 500 and 600 K, then growing and converging to a linear dependence on $T_O$ for large temperatures. The COP of CNTs is shown to remain about the same and approximately equal to 8. These results are shown to be weakly depend on CNT diameter and chirality but not on length. The isothermal entropy change of the CNTs due to the eC is also estimated and shown to depend non-linearly on $T_O$ values. These results predict that CNTs can be considered versatile nanoscale solid refrigerators able to efficiently work over a large temperature span.

**1. Introduction**

Concerns regarding the reduction of greenhouse gas emissions and the increasing demand for cooling have motivated intense research on the development of alternative methods to provide environment-friendly cooling devices [1-12]. One of these alternatives is based on the caloric effect, i. e., the change of the temperature of an object when it is subjected to change of an external field [1,2]. The external field can be of electric, magnetic or elastic/mechanical nature, or a combination of two or more of them. An important advantage of using the caloric effect to develop new cooling devices is that the refrigerant could be solid instead of vapor, what is beneficial for the reduction of greenhouse gases. A disadvantage, however, is that the vapor-compression industry is so well-developed that its refrigerators are not only reliable but also reach efficiencies as high as 60% of Carnot one [3,12,13]. Solutions for the emission of greenhouse gases by vapor-compression refrigerators are also being thought in terms of developing new environment friend fluids or gases [10].

Nevertheless, another advantage of considering caloric effects of solid refrigerants is the possibility to scale down the system and/or applications [14,15]. Vapor-compression technologies cannot be scaled down [8,14,15]. In particular, amongst the caloric effects, the elastocaloric effect (eC) has been considered a promising efficient, non-toxic and noise-free cooling method [15,16], with a predicted efficiency reaching values as large as 87% of Carnot efficiency [13]. eC is the caloric effect under the stimulus of applied strains to the material. Examples of known promising materials for the eC are Ni-Ti [7,13] and Cu-Zn-Al [16,17] shape memory alloys, $Ni_{57}Mn_{18}Ga_{21}In_4$ [18] and CoVGa [19] alloys, $CaF_2$ [20] and natural rubber [21,22].

In view of the advantages of the eC, some computational studies have been performed to predict the eC and the coefficient of performance (COP) of some structures at nanoscale. Lisenkov *et al*. [23] and Zhang [24] were the first to estimate the eC in carbon and boron-nitride (BN) nanotubes, respectively. They have found that the eC values of carbon (BN) nanotubes are up to 30 K (65 K) for 3% (9%) strain. Cantuario and Fonseca [25] were the first to estimate the COP of carbon nanotubes (CNTs) used as eC-based solid refrigerators. They simulated the refrigerator under Otto-like thermodynamic cycles operating at room temperature. The COP is defined as the ratio of the cooling to input power. Cantuario and Fonseca showed that CNTs present COP values between 4 and 6, that is comparable to the COP values of Ni-Ti and other shape memory alloys [7,17]. Other studies have shown the effects of adhesion on the eC of graphene [26], and the eC in ZnO nanowires [27]. Recent caloric studies on nanostructures are barocaloric [28,29] and electrocaloric effects [30].

Most of the above studies predict the eC at operating temperatures ($T_O$s) around 300 K. Few ones have studied the eC and its efficiency at certain $T_O$ windows [16,31,32]. Sehitoglu *et al*. [31] and Chen *et al*. [32] predicted a maximum $T_O$ window of eC in some Ni-Ti based shape memory alloys of about 100 K, while Mañosa *et al*. [16] have got a little bit wider temperature window of 130 K. These values are limited by the fact that the eC in these materials comes from phase transition changes. The eC in CNTs and BN nanotubes was shown not to depend on any phase transition [23,24]. Therefore, one might try to investigate the eC and its efficiency in these nanostructures at different values of $T_O$. In this work, the results for the eC and COP of CNTs as functions of $T_O$, from 300 to 2000 K, are presented. The dependence of these properties on CNT length and chirality is also investigated. We will see that the eC, represented by the change, $\Delta T$, in temperature due to a maximum of 10% of applied strain, has a minimum at $T_O$ values between 500 and 600 K, then grows towards a linear dependence on $T_O$, for large $T_O$, while maintaining roughly the same performance. The isothermal entropy change, $\Delta S$, is also estimated and shown to grow, then, converge to a certain value with increasing $T_O$. These results do not depend on CNT size, but weakly on the CNT chirality.

Section II presents the structure models, the theory behind the eC and the computational methods to find $\Delta T$, $\Delta S$ and COP. Section III presents the results and discussion. Section IV summarizes the main results of this study.

## II. Structure models, eC theory and computational methods.

The following CNTs were considered in the present study: (6,6), (7,4), (8,8), (10,0), (9,7) and (14,0), where (*n*,*m*) are the usual chiral parameters of CNTs. (8,8) and (10,0) CNTs were chosen to compare the results with previous calculations [23,25]. The pairs of tubes (6,6) and (7,4), and (9,7) and (14,0) were chosen because their diameters are approximately equal to those of (8,8) and (10,0) CNTs, respectively ($d_{(6,6)} \approx d_{(7,4)} \approx d_{(10,0)} = 10 \frac{a}{\pi}$, and $d_{(8,8)} \approx d_{(9,7)} \approx d_{(14,0)} = 14 \frac{a}{\pi}$, with $a = a_{cc}\sqrt{3}$, and $a_{cc}$ being the carbon-carbon bond distance). The test of the dependence of the eC on CNT chirality will be done for tubes having the same diameter and length. Two length sizes were analyzed: 10 and 50 nm.

In theory, the eC effect can be represented by two parameters, the isothermal change of the entropy, $\Delta S_{ISO}$; and/or the adiabatic change of temperature, $\Delta T_{ADI}$, of the material due to the variation of the strain, $\varepsilon$. The entropy of an eC material can be written in terms of temperature, $T$, and the applied strain, $\varepsilon$, so the following thermodynamic equation can be written [33]:

$$dS(T,\varepsilon) = \left(\frac{\partial S}{\partial \varepsilon}\right)_T d\varepsilon + \left(\frac{\partial S}{\partial T}\right)_\varepsilon dT . \tag{1}$$

Using the Maxwell relation $\left(\frac{\partial S}{\partial \varepsilon}\right)_T = \left(\frac{\partial \sigma}{\partial T}\right)_\varepsilon$, where $\sigma$ is the applied stress on the eC material, and considering $dT = 0$ ($dS = 0$) for the isothermal (adiabatic) application of strain, $\Delta S_{ISO}$ and $\Delta T_{ADI}$ can, respectively, be obtained from appropriate integrations of eq. (1):

$$\Delta S_{ISO} = \int_{\varepsilon_0}^{\varepsilon_F} \left(\frac{\partial \sigma}{\partial T}\right)_\varepsilon d\varepsilon , \tag{2}$$

$$\Delta T_{ADI} = -\int_{\varepsilon_0}^{\varepsilon_F} \frac{T}{C}\left(\frac{\partial \sigma}{\partial T}\right)_\varepsilon d\varepsilon , \tag{3}$$

where $\varepsilon_0$ and $\varepsilon_F$ are initial and final strains during the corresponding thermodynamic process, and $C$ is the heat capacity at constant strain $\varepsilon$, $C = \left(\frac{dQ}{dT}\right)_\varepsilon = T\left(\frac{dS}{dT}\right)_\varepsilon$. Using the heat capacity of the material, the above-mentioned Maxwell relation and $\left(\frac{\partial S}{\partial \varepsilon}\right)_T d\varepsilon = -\left(\frac{\partial S}{\partial T}\right)_\varepsilon dT$ for the adiabatic process (from $dS = 0$ in eq. (1)), eq. (2) can be integrated to obtain:

$$\Delta S_{\text{ISO}} = -\int_{T_O}^{T_F} \frac{C}{T} dT \cong C \ln\left(\frac{T_O}{T_F}\right), \tag{4}$$

where $T_O$ is the operating temperature (initial in the process), $T_F$ is the temperature at the final applied adiabatic strain $\varepsilon_F$, and the heat capacity was considered to be approximately constant in the temperature interval, $\Delta T_{\text{ADI}} = T_F - T_O$. Eq. (4) can be further simplified to relate $\Delta S_{\text{ISO}}$ to $\Delta T_{\text{ADI}}$ by expanding it up to second order in $\Delta T_{\text{ADI}}/T_O$:

$$\Delta S_{\text{ISO}} \cong -C\left(\frac{\Delta T_{\text{ADI}}}{T_O}\right)\left(1 - \frac{\Delta T_{\text{ADI}}}{2T_O}\right). \tag{5}$$

Both $\Delta S_{\text{ISO}}$ and $\Delta T_{\text{ADI}}$ quantities are not trivial to measure or calculate, so we are going to characterize the eC effect in CNTs by "measuring" $\Delta T_{\text{ADI}}$ through computational numerical experiments and using eq. (5) to estimate the corresponding $\Delta S_{\text{ISO}}$. The CNTs eC and efficiency will be investigated considering an Otto-like thermodynamic cycle consisting of {1 → 2} an adiabatic tensile strain process, departing from the CNT previously equilibrated at a given operating temperature, $T_O$; followed by {2 → 3} an isochoric heat exchange process of the tensioned CNT with a thermal bath at $T_O$; then {3 → 4} application of another adiabatic, now tensile-release, strain process, to bring back the CNT to the initial strain; and, to complete the cycle, the application of {4 → 1} another isochoric heat exchange process with a thermal bath at temperature $T_O$. Figure **1** illustrate the form of the otto-like thermodynamics cycle in terms of the pressure-volume (pV) diagram, with the processes above numbered from 1 to 4 according to the above description. Otto cycles have been considered in previous eC studies [25,32].

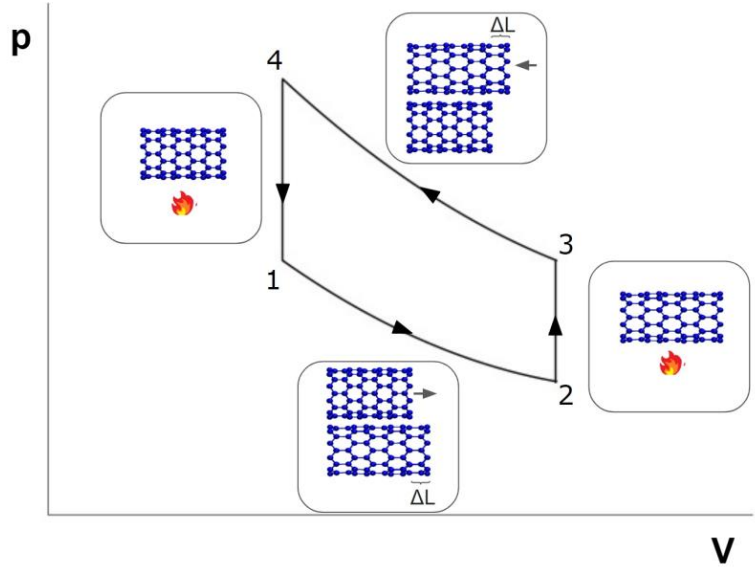

Figure 1: Thermodynamic cycle formed by two processes of adiabatic expansion, $\{1 \to 2\}$ and $\{3 \to 4\}$, and two heat exchange processes, $\{2 \to 3\}$ and $\{4 \to 1\}$, with fixed volume in a thermal reservoir at a certain $T_\text{O}$. $\Delta L$ is the amount of tension strain applied to the nanotube during the adiabatic expansion.

The efficiency of the cycle is computed in terms of the COP, defined as:

$$\text{COP} = \frac{Q_\text{F}}{W_\text{C}}, \qquad (6)$$

where $Q_\text{F}$ and $W_\text{C}$ are the heat exchanged by the refrigerant and the region to be cooled, and the total work on one cycle, respectively. The first can be calculated by

$$Q_\text{F} = mC\Delta T_\text{ADI}, \qquad (7)$$

where $m$ and $C$ are the mass and specific heat capacity of the CNT, and $\Delta T_{ADI}$ is the change in temperature due to the adiabatic application of tensile strain to the structure. Eq. (3) will not be used to obtain $\Delta T_{ADI}$ because although the stress, $\sigma$, on the CNTs can be inferred from molecular dynamics (MD) simulations, its definition depends on the ill-defined cross-sectional area of the CNTs. As the temperature can be directly obtained from the simulations, $\Delta T_{ADI}$ will be obtained directly from those. From now on, we will omit the subscript "ADI" from "$\Delta T$" as it will always correspond to the eC adiabatic change of temperature. However, we have to distinguish the real $\Delta T$, herein called $\Delta T_{REAL}$, from the $\Delta T$ obtained from the computational simulations, herein called $\Delta T_{MD}$, because the classical MD simulations predict values of heat capacity that follows the Dulong-Petit law, while quantum mechanical effects are significant at certain values of temperature. In order to consider the real heat capacity, $C_{REAL}$, of the CNTs that obey the Debye theory, a relation between $\Delta T_{REAL}$ and $\Delta T_{MD}$ should be found. It is, in fact, simple and comes from the exchanged heat with the thermal bath during the isochoric process that is given by $Q_F = mC_{REAL}\Delta T_{REAL} = mC_{MD}\Delta T_{MD}$. So,

$$\Delta T_{REAL} = \left(\frac{C_{MD}}{C_{REAL}}\right)\Delta T_{MD}. \tag{8}$$

In order to obtain $\Delta T_{REAL}$, we need to obtain the values of $C_{REAL}$ for CNTs, for every temperature. Experimental values of $C_{REAL}$ for CNTs are only available at temperatures smaller than or equal to 300 K [34]. As the Debye temperature of carbon structures is larger than 300 K [35], it is important to find out a way to estimate the $C_{REAL}$ for CNTs at temperature values within the range from 300 to 2000 K. It can be done using, for example, the theoretical approach developed by Mir *et al*. [36]. Based on the Debye's formula, the heat capacity of a solid can be written as [36,37]:

$$C_{\text{REAL}}(T) = C^{\text{TE}} \left(\frac{T}{T_{\text{D}}}\right)^3 \int_0^{T_{\text{D}}/T} \frac{x^4 e^x}{(e^x-1)^2} dx, \qquad (9)$$

where $T_{\text{D}}$ is the Debye temperature given by $T_{\text{D}} = h\nu_{\text{D}}/k_{\text{B}}$, with $h$ and $k_{\text{B}}$ being the Planck and Boltzmann constants, respectively, and $\nu_D$ the Debye frequency of the material that, for carbon nanotubes, is equal to $4.3867 \times 10^{13}$ s$^{-1}$ [36]. It gives $T_{\text{D}} \cong 2107$ K. The value of $C^{\text{TE}}$ can be obtained using the eq. (9) with the known value of the $C_{\text{REAL}} = 645$ J kg$^{-1}$ K$^{-1}$ at $T = 300$ K as considered in previous studies [23,34]. We found $C^{\text{TE}} = 10216.4$ J kg$^{-1}$ K$^{-1}$, that can be, then, used to calculate $C_{\text{REAL}}$ for the values of operating temperatures in the range from 400 to 2000 K. Table **I** below, shows the values of $C_{\text{REAL}}$ obtained from the above method. Section **S1** of Supplemental Material (SM) [38] shows a graphic of $C_{\text{REAL}}(T)$ for $0 < T < 3000$ K, using eq. (9) and the above value of $C^{\text{TE}}$ in order to verify the Debye curve for CNTs.

Table **I**: Values of the real heat capacity, $C_{\text{REAL}}$, of CNTs for different operating temperature values, $T_{\text{O}}$, considered in this work, calculated by eq. (9) and the method described in the text.

| $T_{\text{O}}$ [K] | 300 | 400 | 500 | 600 | 700 | 800 |
|---|---|---|---|---|---|---|
| $C_{\text{REAL}}$ [J kg$^{-1}$ K$^{-1}$] | 645 | 1151.3 | 1607.0 | 1971.8 | 2252.2 | 2465.8 |
| $T_{\text{O}}$ [K] | 900 | 1000 | 1100 | 1200 | 1300 | 1400 |
| $C_{\text{REAL}}$ [J kg$^{-1}$ K$^{-1}$] | 2629.4 | 2756.1 | 2855.6 | 2934.7 | 2998.5 | 3050.5 |
| $T_{\text{O}}$ [K] | 1500 | 1600 | 1700 | 1800 | 1900 | 2000 |
| $C_{\text{REAL}}$ [J kg$^{-1}$ K$^{-1}$] | 3093.4 | 3129.2 | 3159.3 | 3184.8 | 3206.7 | 3225.5 |

The total work on one thermodynamic cycle of the eC process, $W_{\text{C}}$, can be calculated as follows. As the adiabatic processes of the cycle are performed without heat exchange between the system and the exterior media, the total energy before, $E_{\text{B}}$, and after, $E_{\text{A}}$, the application of tension or tension-release to the CNT can be computed and the work done or received by the CNT can be simply calculated as

$$W_C = (E_A - E_B)_{\text{TENSION}} + (E_A - E_B)_{\text{TENSION-RELEASE}}. \tag{10}$$

No work is done or received by the CNT under the heat exchange processes {2 → 3} and {4 → 1}, because its size is kept fixed during them. When applying tension to the CNT, it suffers the work from the external media, so its total energy increases, or $(E_A - E_B)_{\text{TENSION}} > 0$. When allowing the tension to be released, the CNT energy decreases, or $(E_A - E_B)_{\text{TENSION-RELEASE}} < 0$. The sum of these terms, then, represents the net work performed on the CNT during one thermodynamic cycle.

The protocols of MD simulations are given as follows. They were performed with the software LAMMPS (Large-scale Atomic/Molecular Massively Parallel Simulator) [39]. The Adaptive Intermolecular (AI) Reactive Empirical Bond Order (REBO) [40,41] classical force field was used to simulate the carbon-carbon interactions. AI+REBO is a well-known potential used to simulate structural, mechanical and thermal properties of carbon nanostructures [42-51]. Besides, it has been used before to obtain the eC properties of CNTs and graphene [23,25,26].

The structure of all CNTs studied here was first optimized through energy minimization algorithms and, then, equilibrated at $T_O$ before starting the thermodynamic cycle previously described. The energy minimizations were performed imposing periodic boundary conditions (PBC) along the CNT axis including the possibility to relax the size along the PBC direction. We followed a protocol of energy minimizations that ensures finding out the structure of smallest energy by performing combinations of energy minimization and free evolution algorithms as suggested by Sihn *et al*. [52] and recently implemented by Kanegae and Fonseca [53] in a study of graphynes. Thermal equilibration of the CNTs is performed by fixing the carbon atoms at the extremities and then applying a Langevin thermostat to all other atoms with damping factor of 100 fs and timestep of 0.5 fs, for a total time of 60 ps. Although previously energy-minimized, performing thermal equilibration on a CNT structure with fixed extremities might generate internal thermal stresses due to thermal expansion. However, the REBO potential predicts CNT axial linear thermal expansion coefficients of the order of ~ $10^{-6}$ K$^{-1}$ [45,54-56], from negative to positive values in a wide range of temperatures. So, in the temperature span of about 2000 K, the thermal strain will be maximum about ~ $10^{-3}$, or ~ 0,1%, that is much less than the 10% of maximum strain considered in this study. We will, then, neglect the thermal expansion effects.

The adiabatic processes were simulated by keeping one end of the CNT fixed and applying a constant speed to the atoms of the other end. The atoms between the extremities have the initial velocities set to be compatible with a Boltzmann distribution corresponding to the thermal equilibrium at the operating temperature, $T_O$. They are not thermostated during the adiabatic processes. Fixing the speed of the moving carbon atoms at one end of the CNT and the total simulation time, fixes the strain rate of the tensile or tensile-release processes. In order to obtain a better precision on the eC effect during the tensile strain simulations, the CNTs are strained using a timestep much smaller than that of usual simulations of carbon nanostructures using the AIREBO potential. The value of 0.02 fs of timestep (25 times smaller than the usual 0.5 fs of timestep) was used by a total amount of time of 32 ps (or 1600000 total simulation steps) to give 10%/32 ps or 0.003125 ps$^{-1}$ of strain rate, the same value used in previous studies [23,25]. As we shall see in the next section, the strain rate can affect the results for CNTs of different lengths. In order to keep the same strain rate and timestep, longer the CNT, smaller the chosen speed of movement of the atoms of its non-fixed extremity and longer the total simulation time. The reason is that longer the CNT, larger the time needed for the atoms far from those being pulled to feel the strain and equilibrate. In the heat exchange, the atoms of both ends of the CNT are kept fixed and the thermal equilibrium simulation of the remaining atoms are performed at $T_O$ using Langevin thermostat with damping factor of 100 fs and timestep of 0.5 fs, for a total time of 60 ps.

The eC temperature variation, $\Delta T_{MD}$, from the MD simulations, will be computed as the difference between $T_O$ and the last point of a moving average (over 1000 points) curve of temperature versus time collected along the adiabatic tensile simulations of the CNTs every 2 fs (or every 100 steps, which gives 16000 data points). As explained above, the eC corresponding to the real temperature change, $\Delta T_{REAL}$, is obtained using eq. (8) and the values of $C_{REAL}$ showed in Table **I** for each value of the operating temperature, $T_O$. The MD values of the heat capacity, $C_{MD}$, are obtained from running previously series of MD simulations with each CNT, each simulation at a different value of temperature ranging from 300 to 2000 K in steps of 50 K. From the slope of the energy versus temperature of each CNT obtained from these simulations, $C_{MD}$ is determined. Each simulation is run by 60 ps with fixed extremities.

For the calculation of the COP, it is enough to use the value of $\Delta T_{MD}$ because $Q = mC_{MD}\Delta T_{MD}$. The work per cycle, $W_C$, is calculated by eq. (10) where $E_A$ and $E_B$ are taken from the MD simulations.

## III. Results and discussion

In this section, the results for the eC, $\Delta T$ and $\Delta S$, and COP of all CNTs for a range of operating temperatures, $T_O$, from 300 K to 2000 K are presented, including the dependence of these properties on the CNT size. The results for all CNTs of 10 nm length are presented first, including the analysis of the dependence on the CNT chirality and diameter. The length dependence of the eC and COP of CNTs is presented later.

Preliminary results of the $C_{MD}$ of all CNTs are shown in Table **II**. The curves of energy versus equilibrium temperature of all CNTs from which the data shown in Table **II** were obtained are shown in Section **S2** of the Supplemental Materials [38].

Table **II**: Values of the MD heat capacity, $C_{MD}$ [J kg$^{-1}$ K$^{-1}$], of CNTs from the MD simulations.

| (6,6) | (7,4) | (8,8) | (10,0) | (9,7) | (14,0) |
|---|---|---|---|---|---|
| 2065,07 | 1987,82 | 2051,51 | 2072,43 | 2056,94 | 2057,84 |

### III.1 eC and COP of CNTs

Figure **2** shows the eC, $\Delta T_{REAL}$, and $\Delta S$ (per kg) of all CNTs of 10 nm length studied here, subjected to the thermodynamic refrigerator cycle described in the previous section, running at operating temperatures ranging from 300 to 2000 K. Quantum mechanical effects on the heat capacity of CNTs affect the dependence of $\Delta T_{REAL}$ on $T_O$. It presents a minimum in the considered operating temperature interval and after that, it grows towards a linear dependence on $T_O$ when $T_O \to \infty$. As the heat capacity of a system tends to a constant according to the Dulong-Petit law, the linear growth of $\Delta T_{REAL}$ reflects this limit. Fig. **S3.1** in Section **S3** of the SM [38]

shows the linear dependence of the values of $\Delta T_{MD}$ on $T_O$. As the results for $\Delta T_{MD}$ reflects only the classical behavior of the heat capacity of CNTs, the linear growth of $\Delta T_{REAL}$ on $T_O$, for large $T_O$, represents the Dulong-Petit law when $T_O \to \infty$.

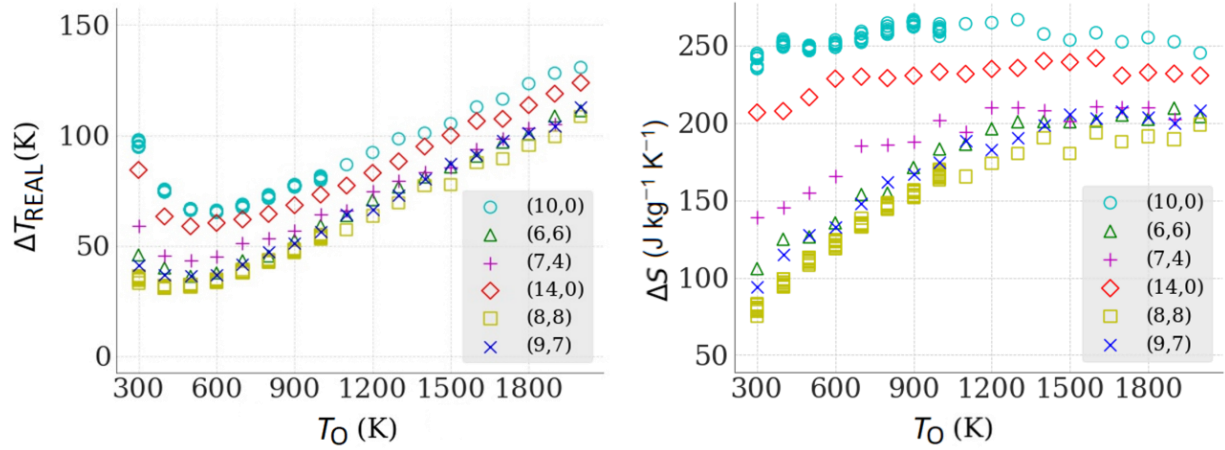

Figure 2: Adiabatic temperature change, $\Delta T_{REAL}$ (left panel), and the specific (per kg) isothermal change of entropy, $\Delta S$ (right panel), versus operating temperature, $T_O$, of all CNTs studied here with 10 nm length. For (8,8) and (10,0) CNTs, results from the simulations of 10 different microcanonical states corresponding to the same thermodynamic macroscopic states (same $T_O$, zero pressure and size) are superposed.

In the seminal study by Cantuario and Fonseca [25], fluctuations of the eC and COP of CNTs have not been estimated. In the present work, for the (8,8) and (10,0) CNTs, they were calculated for a subrange of operating temperatures from 300 to 1000 K. 10 different microcanonical states corresponding to the same thermodynamic macroscopic states (same $T_O$, zero pressure and size) were considered and subjected to the same thermodynamic cycle. Different microcanonical states can be computationally obtained by using different values of *seed* in the algorithms that generate the initial distribution of velocities corresponding to the $T_O$ temperature, or simulate the thermal bath. The idea is to estimate the fluctuations in the eC and COP of the CNTs under the same values of $T_O$, in order to further verify if they depend on different $T_O$s. In figure **2**, the superposed circle and squared points for each temperature in the interval from 300 to 1000 K, correspond to the simulated eC of 10 different microcanonical states of the (10,0) and (8,8) CNTs. We can clearly see that the deviations from the average values are relatively small for both the eC $\Delta T_{REAL}$ and $\Delta S$. We, then, decided to extend these calculations for $T_O$ up to 2000 K without repeating for different microcanonical states. Also, we decided to not repeat the simulations for different microcanonical states to obtain the eC of the other CNTs.

As figure **2** shows, $\Delta T_{REAL}$ of the zigzag CNTs are few tens of Kelvin larger than that of other CNTs. Besides, smaller the $T_O$, larger this difference in $\Delta T_{REAL}$. It might be a consequence of differences in the isothermal entropy changes among different nanotubes. Panel right of figure **2** presents the results for $\Delta S$ of all CNTs and it is possible to see that the isothermal entropy change in zigzag CNTs are larger than that of the other CNTs. The entropy change is roughly constant for the zigzag CNTs while for the other CNTs, it increases with $T_O$ and roughly converges after $T_O \sim 1000$ K.

Figure **2** also allows to observe the weak dependence of the eC on the CNT diameter. When comparing the eC of CNTs of same type of chirality, we notice that larger the diameter, smaller the values of $\Delta T_{REAL}$ and $\Delta S$. Figure **S3.2** in SM [38] shows the values of $\Delta T_{MD}$ for each pair of tubes of same chirality but different diameters.

Figure **3** shows the COP of the (10,0) and (8,8) CNTs for each $T_O$ value in the full range from 300 to 2000 K. Here, as in figure **2**, for the subrange from 300 to 1000 K, the results for the COP are shown for every simulation with a statistically different microcanonical state. Differently from the eC values, the COP of the CNTs visibly fluctuates with different microcanonical states, at least with large magnitude for the (8,8) CNT. However, the averages of these COP values at each $T_O$ from 300 to 1000 K are not so different from those that were calculated at larger temperatures. This suggests that the COP of CNTs might not depend on the operating temperature of the refrigerator. In order to verify the generality of this result, we plotted together and showed in figure **4**, the COP of all CNTs studied here at the full $T_O$ range.

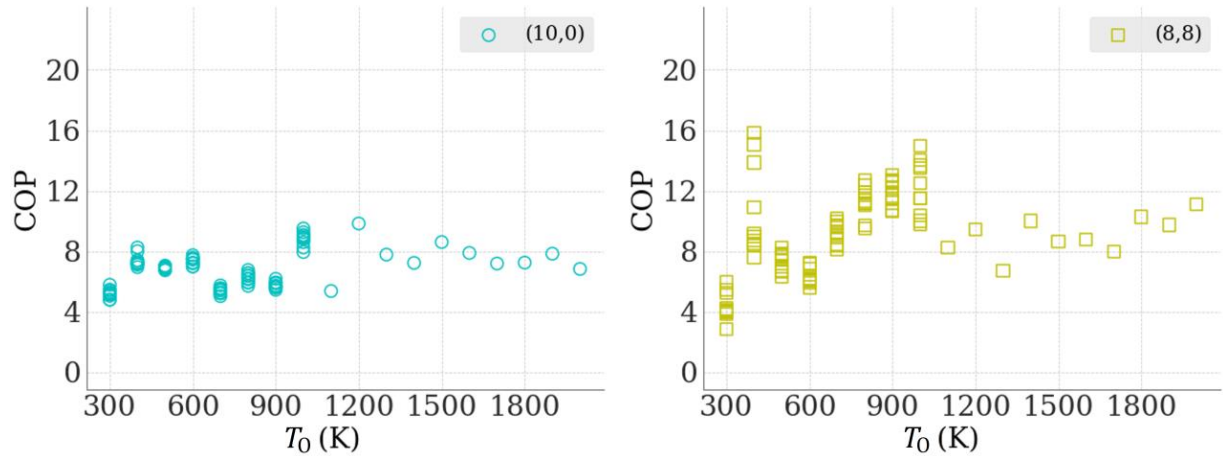

Figure **3**: COP of the (10,0) and (8,8) CNTs for each $T_O$ for the full range from 300 to 2000 K. For the subrange from 300 to 1000 K, the results for the COP of the (10,0) and (8,8) CNTs corresponding to each of the 10 different microcanonical states are shown.

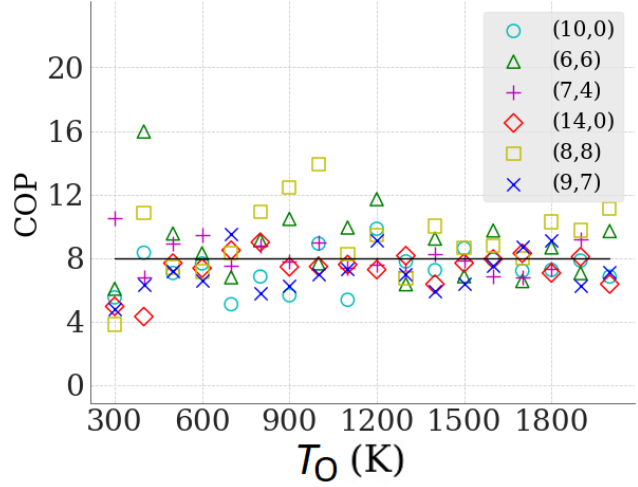

Figure **4**: COP of all CNTs of 10 nm length studied here for each $T_O$ from 300 to 2000 K. For the (8,8) and (10,0) CNTs, the values of COP shown here at the subrange from 300 to 1000 K were averaged over the simulations of 10 different microcanonical states. The horizontal line represents the average of all values.

Figure **4** allows us to conclude that although the COP of a refrigerator using the eC of the CNTs fluctuates significantly, it is not possible to identify any special dependence on the operating temperature, $T_O$. This is, in fact, a good result because it indicates that CNTs can be valuable solid refrigerators with good efficiency running at a large temperature span. Most of the simulations resulted in COP > 4.

In order to obtain the COP, it is necessary to calculate the work received by the CNT during one thermodynamical cycle, $W_C$. In figure **S4** of the SM [38], we show two examples of how the total energy of two different CNTs varies with time during their adiabatic expansion and contraction, at two different $T_O$ values.

## III.2 Length dependence of the eC and COP of CNTs

The eC and COP of (10,0) and (8,8) CNTs of 50 nm length are presented in this subsection. The same thermodynamic cycle simulated with all 10 nm length CNTs was applied to the 50 nm ones, but for temperature values only between 300 and 1000 K. The simulations were performed with the same parameters and conditions as previously, except for the total simulation times of the adiabatic processes that were changed in order to verify the effect of the strain rate on the results.

The first tests consisted of identifying the largest strain rate for which the eC results become independent on the rate.

Table **III**: Values of the eC $\Delta T$, $\Delta S$ and COP of (10,0) CNT of 50 nm length, simulated at $T_O$ = 300 K, for different values of the total time of the adiabatic tension and tension-release simulations. The timestep and amount of strain of these simulations were all the same and equal to 0.02 fs and 10%, respectively, so the differences are in the strain rate.

| Simulation time [ps] | Strain rate [nm ps$^{-1}$] | $\Delta T_{MD}$ [K] | $\Delta T_{REAL}$ [K] | $\Delta S$ [J kg$^{-1}$ K$^{-1}$] | COP |
|---|---|---|---|---|---|
| 32 | 0.15625 | 28.39 | 91.22 | 225.81 | 1.89 |
| 64 | 0.078125 | 29.37 | 94.37 | 234.67 | 3.60 |
| 96 | 0.0520833 | 31.36 | 100.76 | 252.87 | 4.60 |
| 128 | 0.0390625 | 31.59 | 101.50 | 254.99 | 5.71 |
| 160 | 0.03125 | 31.57 | 101.44 | 254.82 | 6.87 |
| 200 | 0.025 | 29.76 | 95.62 | 238.22 | 5.72 |
| 250 | 0.02 | 30.17 | 96.94 | 241.95 | 6.09 |
| 320 | 0.015625 | 31.97 | 102.72 | 258.51 | 6.72 |

Table **III** shows the values for $\Delta T$, $\Delta S$ and COP of 50 nm (10,0) CNTs obtained from simulations of the full thermodynamical cycle at $T_O$ = 300 K, with the same parameters and conditions as done in the previous section (i.e., 5 nm or 10% of total tensile strain and timestep = 0.02 fs), but for different strain rates represented by the values of the time of the tensile strain and tensile-release strain simulations. While $\Delta T$ and $\Delta S$ did not change significantly, the COP increases with the total simulation times up to 128 ps. For times equal or larger than 128 ps, the COP converges. This result might reflect the effect of the strain rate that becomes more relevant for long tubes and same tensile strain simulation times considered for short tubes. Larger the strain rate, lesser the time the system has to equilibrate. For 10 nm length CNT, the total strain of 10% (or $\Delta L$ = 1 nm) was simulated in 32 ps, what corresponds to a tensile strain speed of 0.03125 nm/ps, or a strain rate of 10/32 = 0.3125 % ps$^{-1}$. For the strain rate on a 50 nm length CNT to be the same as that of simulating 10% strain of a 10 nm length CNT, the total simulation time should be increased proportionally. As 10% of 50 nm is 5 nm, we must tensile strain the structure during 5 nm / 1nm times 32 ps or 160 ps (5 nm/160 ps = 1 nm/32 ps = 0.03125 nm/ps). As seen in Table **III**, from 128 ps of total simulation time, the COP reaches a value that roughly converges. Therefore, the value of 5/128 = 0.0390625 nm/ps might be considered an estimate for the largest strain rate able to allow the CNT structure to equilibrate during the eC. Here, we decided to set the value of strain rate of the simulations of 50 nm CNTs to that used in the simulations of 10 nm long CNTs (strain rate of 0.3125 % ps$^{-1}$). It means that the total simulation time of the tensile strain and tensile strain release processes, to obtain the eC of the 50 nm length CNTs at other values of $T_O$, is set to 160 ps.

Table **IV**: Values of the COP of (10,0) and (8,8) CNTs of 10 and 50 nm lengths for each $T_O$ from 300 to 1000 K, for comparison.

| CNT | $L$ [nm] | 300 K | 400 K | 500 K | 600 K | 700 K | 800 K | 900 K | 1000 K | Average |
|---|---|---|---|---|---|---|---|---|---|---|
| (8,8) | 10 | 4.35 | 10.70 | 7.34 | 6.43 | 9.22 | 11.25 | 11.88 | 12.18 | 9.17 |
|  | 50 | 8.07 | 9.52 | 9.20 | 9.15 | 8.29 | 10.29 | 8.66 | 8.07 | 8.91 |
| (10,0) | 10 | 5.20 | 7.38 | 6.90 | 7.33 | 5.36 | 6.24 | 5.75 | 8.82 | 6.62 |
|  | 50 | 6.87 | 5.73 | 6.65 | 6.74 | 5.61 | 5.90 | 4.45 | 5.41 | 5.92 |

Table **V**: Values of $\Delta T_{REAL}$ [K] of (10,0) and (8,8) CNTs of 10 and 50 nm lengths for each $T_O$ from 300 to 1000 K, for comparison.

| CNT | $L$ [nm] | 300 K | 400 K | 500 K | 600 K | 700 K | 800 K | 900 K | 1000 K |
|---|---|---|---|---|---|---|---|---|---|
| (8,8) | 10 | 35.12 | 31.26 | 32.00 | 34.12 | 38.19 | 43.04 | 47.72 | 54.08 |
|  | 50 | 37.47 | 34.25 | 34.32 | 36.72 | 41.27 | 44.74 | 50.21 | 55.10 |
| (10,0) | 10 | 96.30 | 74.82 | 66.38 | 65.48 | 67.94 | 72.05 | 77.13 | 80.53 |
|  | 50 | 101.44 | 74.74 | 70.70 | 69.36 | 70.03 | 73.12 | 73.06 | 80.50 |

Tables **IV** and **V** show the COP and $\Delta T_{REAL}$ of 10 and 50 nm length (8,8) and (10,0) CNTs, for $T_O$ ranging from 300 and 1000 K. We note that the COPs for the longer CNTs, in average, remained the same as that of 10 nm length CNTs. The values of $\Delta T_{REAL}$ are about the same too, i.e., there is no sign of length dependence of the eC in CNTs.

**IV. Conclusions**

The eC of CNTs of different lengths, chiralities and diameters were obtained from MD simulations for a large temperature span. We show that while the COP of the CNTs as a solid refrigerant remains roughly the same for the range 300 K ≤ $T_O$ ≤ 2000 K considered here, the eC $\Delta T$ and $\Delta S$ change nonlinearly with $T_O$. Classical thermal behavior is observed to occur for $T_O$ > 1000 K since the dependence of $\Delta T$ on $T_O$ becomes linear. The results show a weak dependence of the eC on CNT diameter and no dependence on length. The adiabatic temperature change, $\Delta T$, and the isothermal entropy change, $\Delta S$, of zigzag CNTs during one cycle of the simulated thermodynamic cycle of the refrigerator machine were shown to be larger than that of other chiral CNTs, mostly at low $T_O$.

Different from other good eC materials, the eC in CNTs does not depend on phase transitions. Therefore, CNTs can be explored as solid refrigerators at large ranges of operating temperatures. In particular, our results predict that not only the eC is significant at a large temperature span but also with approximately the same good efficiency with COP values estimated to be about 8. In view of the quest for new, safe and low-cost refrigeration methods, together with the increasing miniaturization trends in technology, we believe the results from the present study will guide and motivate further experiments on the development of nanorefrigerators.

**Acknowledgements**

TNYS and AFF acknowledge support from São Paulo Research Foundation (FAPESP) through grants #2020/05333-1 and #2020/02044-9, respectively. AFF is a fellow if CNPq – Brazil (303284/2021-8). They also acknowledge support from the John David Rogers Computing Center (CCJDR) at the Institute of Physics "Gleb Wataghin", University of Campinas.

Supplementary Material for

# High performance of carbon nanotube refrigerators over a large temperature span

*Tatiana Naomi Yamamoto Silva, Alexandre F. Fonseca* [*]

Applied Physics Department, Institute of Physics "Gleb Wataghin", University of Campinas - UNICAMP, 13083-859, Campinas, São Paulo, Brazil.

**Corresponding Author**

* Phone: +55 19 3521-5364. Email: Alexandre F. Fonseca – afonseca@ifi.unicamp.br

**S1. Dependence on temperature of the heat capacity of carbon nanotubes (CNTs).** Below, it is presented the plot of the Debye's formula for the heat capacity of the CNT, eq. (9) of the main text, as considered by Mir *et al*. [SM1]. The value of $C^{TE} = 10216.4$ J kg$^{-1}$ K$^{-1}$ needed to plot the graphics was obtained using the experimental value of 645 J kg$^{-1}$ K$^{-1}$ at 300 K [SM2].

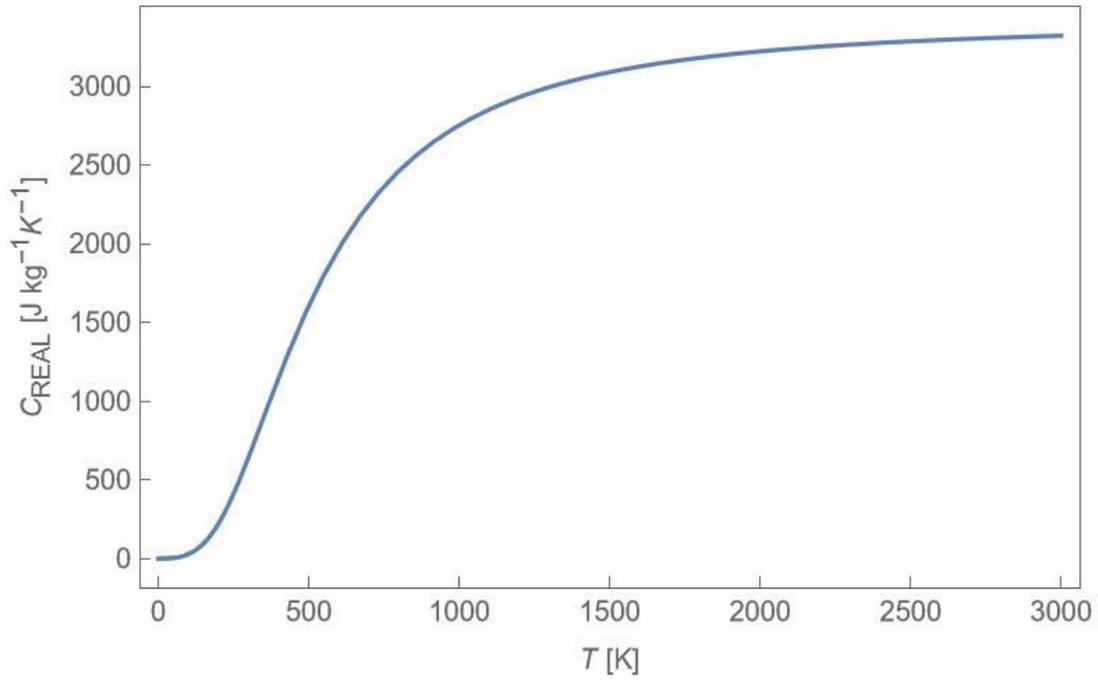

**Figure S1**. Real heat capacity $C_{REAL}(T)$ of CNTs versus temperature $T$ in the range $0 < T < 3000$ K.

**S2. Curves of energy versus equilibrium temperature for determination of the heat capacity from MD simulations.**

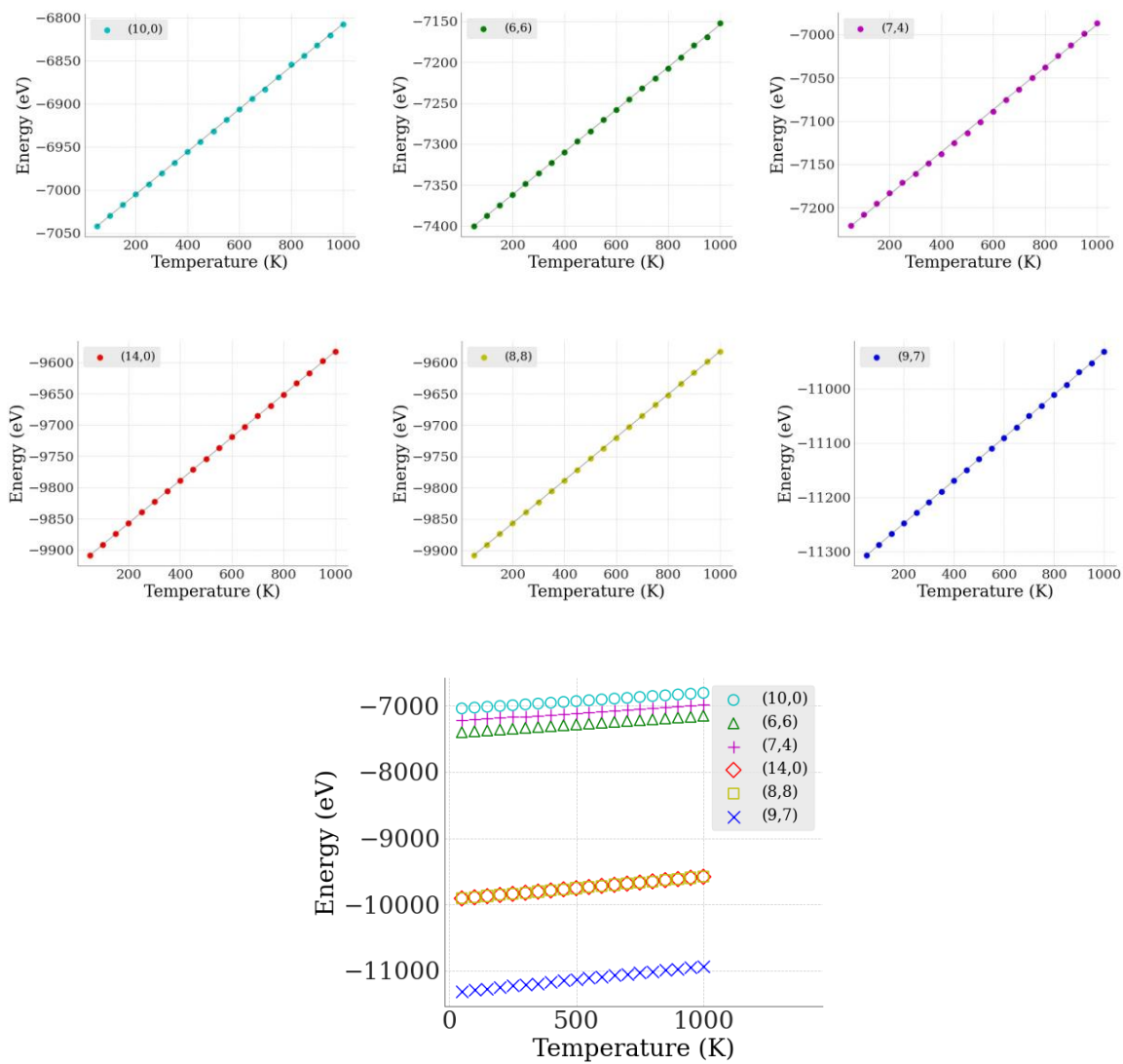

**Figure S2**. Energy versus equilibrium temperature for each CNT studied here. The slope of these curves determines the heat capacity from MD simulations. Lines represent linear fittings of the data. The bottom graphics shows the data all together.

**S3. eC effect, $\Delta T_{MD}$, as obtained directly from the classical MD simulations.** Figs. **S3.1** and **S3.2** show $\Delta T_{MD}$ for all CNTs and $\Delta T_{MD}$ for CNTs of same chirality, respectively. The last figure allows to see the small dependence of the eC on tube diameter.

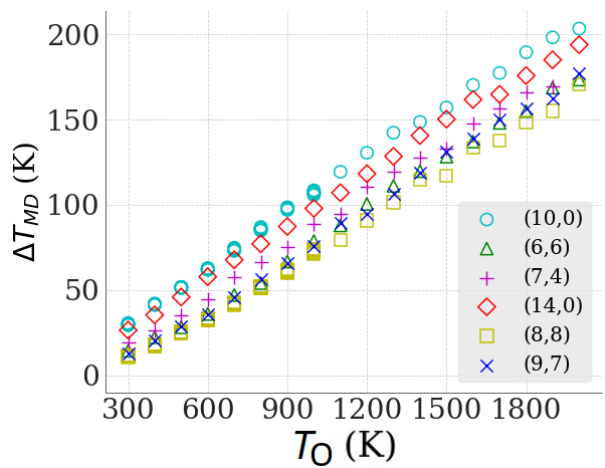

**Figure S3.1** eC, $\Delta T_{MD}$, values obtained for thermodynamic cycles with temperatures between 300 and 2000 K for the CNTs (10,0), (6,6), (7,4), (14,0), (8,8) and (9,7) with 10 nm of length. For the tubes (10,0) and (8,8) and temperatures ranging from 300 to 1000 K, different results corresponding to ten different statistically independent runs are shown by the superposed circles and squares.

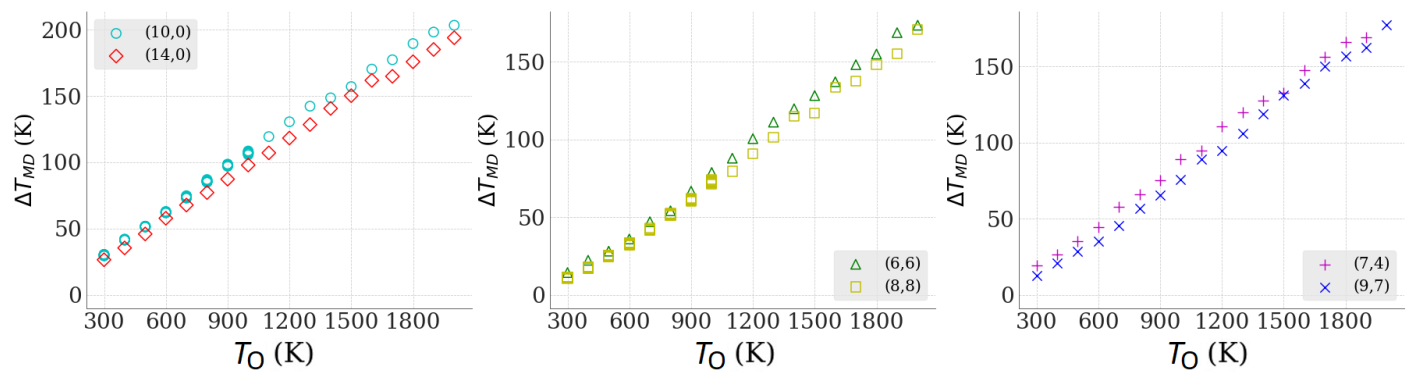

**Figure S3.2** eC, $\Delta T_{MD}$, values obtained for thermodynamic cycles with temperatures between 300 and 2000 K for the CNTs (10,0) and (14,0) (left), (6,6) and (8,8) (middle) and (7,4) and (9,7) (right) with 10 nm of length. In all of them, larger the diameter, smaller $\Delta T_{MD}$.

**S4. Total energy of the CNT under tension and tension-release.** Figure **S4** shows two examples of variation of the total energy with time during the adiabatic expansion and contraction processes of different CNTs and $T_O$s. As there is not any heat exchange, although the atoms thermally vibrate, the total energy simply follows a curve representing the variation of the energy due to the application of the tensile strain. The work done or received by the CNT can be simply calculated as the difference between the total energy after, $E_A$, and total energy before, $E_B$, the process. As can be seen in figure **S4**, it is enough to take the first and last values of the total energy to compute the work done or received by the CNT.

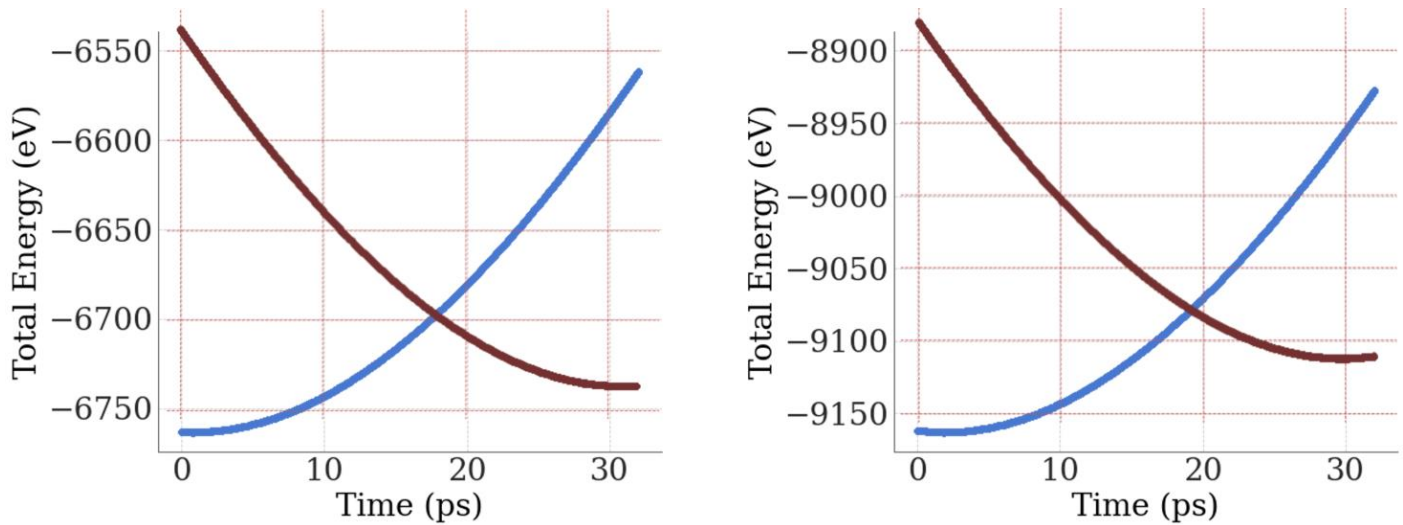

**Figure S4.** Total energy of (10,0) (left panel) and (8,8) (right panel) during the adiabatic expansion (blue lines) and contraction (dark red lines) at $T_O$ = 1000 K (left panel) and 2000 K (right panel).